\documentstyle[aps,prl,multicol,graphics]{revtex}
\begin{document}

\widetext
\begin{multicols}{2}

{\bf Comment on ``Universal relation between the Kolmogorov-Sinai
Entropy and the thermodynamic entropy in simple liquids''}

\vskip 5pt

In an intriguing paper [1], Dzugutov, Aurell and Vulpiani (DAV) establish 
a quantitative connection in the form of a universal linear 
law between the Kolmogorov-Sinai entropy $h_{KS}$ and 
the excess entropy $S_{ex}$ for three spherical 
potentials: $h^{DAV}_{KS}/\Gamma_{coll} = 0.62 + 1.106 S_{ex}$ $(1)$.
Here $\Gamma_{coll}$ is the average pair collision frequency of an 
equivalent hard sphere fluid. We applied this relation to a 
hard sphere fluid, comparing it with the values of $h_{KS}^{DP}$ 
given by Dellago and Posch (DP)
[2]. In Fig.\ 1 we plot $h_{KS}^{DP}$, as well as $(1)$.
In the calculation of the curve labelled DAV we used 
$S_{ex} = 3-12(9-\pi n \sigma^3)/(6-\pi n\sigma^3)^2$ based on the
Carnahan-Starling formula for the pressure of a hard sphere fluid and
on thermodynamic integration [1]. It is clear that (1)
does not apply to the $h_{KS}^{DP}$.  
The same holds for the case of hard disk systems in two
dimensions, for which the assumption of any linear law like
$(1)$, leads to the same contradictions.
\begin{figure}
\setlength{\unitlength}{0.240900pt}
\ifx\plotpoint\undefined\newsavebox{\plotpoint}\fi
\sbox{\plotpoint}{\rule[-0.200pt]{0.400pt}{0.400pt}}%
\begin{picture}(900,594)(0,0)
\font\gnuplot=cmr10 at 10pt
\gnuplot
\sbox{\plotpoint}{\rule[-0.200pt]{0.400pt}{0.400pt}}%
\put(100.0,213.0){\rule[-0.200pt]{4.818pt}{0.400pt}}
\put(80,213){\makebox(0,0)[r]{5}}
\put(859.0,213.0){\rule[-0.200pt]{4.818pt}{0.400pt}}
\put(100.0,344.0){\rule[-0.200pt]{4.818pt}{0.400pt}}
\put(80,344){\makebox(0,0)[r]{10}}
\put(859.0,344.0){\rule[-0.200pt]{4.818pt}{0.400pt}}
\put(100.0,475.0){\rule[-0.200pt]{4.818pt}{0.400pt}}
\put(80,475){\makebox(0,0)[r]{15}}
\put(859.0,475.0){\rule[-0.200pt]{4.818pt}{0.400pt}}
\put(100.0,82.0){\rule[-0.200pt]{0.400pt}{4.818pt}}
\put(100,41){\makebox(0,0){0}}
\put(100.0,534.0){\rule[-0.200pt]{0.400pt}{4.818pt}}
\put(383.0,82.0){\rule[-0.200pt]{0.400pt}{4.818pt}}
\put(383,41){\makebox(0,0){0.4}}
\put(383.0,534.0){\rule[-0.200pt]{0.400pt}{4.818pt}}
\put(667.0,82.0){\rule[-0.200pt]{0.400pt}{4.818pt}}
\put(667,41){\makebox(0,0){0.8}}
\put(667.0,534.0){\rule[-0.200pt]{0.400pt}{4.818pt}}
\put(100.0,82.0){\rule[-0.200pt]{187.661pt}{0.400pt}}
\put(879.0,82.0){\rule[-0.200pt]{0.400pt}{113.705pt}}
\put(100.0,554.0){\rule[-0.200pt]{187.661pt}{0.400pt}}
\put(286,498){\makebox(0,0){$h_{KS}\sqrt{m\sigma^2/NK}$}}
\put(529,21){\makebox(0,0){$n \sigma^3$}}
\put(383,292){\makebox(0,0)[l]{DP}}
\put(560,234){\makebox(0,0)[l]{DAV}}
\put(100.0,82.0){\rule[-0.200pt]{0.400pt}{113.705pt}}
\sbox{\plotpoint}{\rule[-0.500pt]{1.000pt}{1.000pt}}%
\put(100,82){\usebox{\plotpoint}}
\put(100.00,82.00){\usebox{\plotpoint}}
\put(120.60,84.57){\usebox{\plotpoint}}
\put(140.99,88.25){\usebox{\plotpoint}}
\put(161.59,90.82){\usebox{\plotpoint}}
\put(182.00,94.37){\usebox{\plotpoint}}
\put(202.40,98.05){\usebox{\plotpoint}}
\put(222.88,101.22){\usebox{\plotpoint}}
\put(243.22,105.15){\usebox{\plotpoint}}
\put(263.59,108.82){\usebox{\plotpoint}}
\put(283.82,113.35){\usebox{\plotpoint}}
\put(304.07,117.77){\usebox{\plotpoint}}
\put(324.33,122.08){\usebox{\plotpoint}}
\put(344.46,127.12){\usebox{\plotpoint}}
\put(364.60,132.15){\usebox{\plotpoint}}
\put(384.67,137.42){\usebox{\plotpoint}}
\put(404.81,142.45){\usebox{\plotpoint}}
\put(424.64,148.47){\usebox{\plotpoint}}
\put(444.49,154.43){\usebox{\plotpoint}}
\put(464.57,159.64){\usebox{\plotpoint}}
\put(484.42,165.60){\usebox{\plotpoint}}
\put(504.17,171.79){\usebox{\plotpoint}}
\put(524.01,177.75){\usebox{\plotpoint}}
\put(543.86,183.72){\usebox{\plotpoint}}
\put(563.66,189.87){\usebox{\plotpoint}}
\put(583.78,194.95){\usebox{\plotpoint}}
\put(603.92,199.98){\usebox{\plotpoint}}
\put(624.31,203.66){\usebox{\plotpoint}}
\put(644.72,207.22){\usebox{\plotpoint}}
\put(665.38,208.80){\usebox{\plotpoint}}
\put(686.09,208.49){\usebox{\plotpoint}}
\put(706.49,204.88){\usebox{\plotpoint}}
\put(726.39,199.12){\usebox{\plotpoint}}
\put(744.60,189.25){\usebox{\plotpoint}}
\put(760.74,176.23){\usebox{\plotpoint}}
\put(775.50,161.69){\usebox{\plotpoint}}
\put(787.63,144.87){\usebox{\plotpoint}}
\put(798.26,127.05){\usebox{\plotpoint}}
\put(808.10,108.78){\usebox{\plotpoint}}
\put(816.82,89.95){\usebox{\plotpoint}}
\put(820,82){\usebox{\plotpoint}}
\sbox{\plotpoint}{\rule[-0.400pt]{0.800pt}{0.800pt}}%
\put(100,92){\usebox{\plotpoint}}
\put(100,92.34){\rule{1.800pt}{0.800pt}}
\multiput(100.00,90.34)(4.264,4.000){2}{\rule{0.900pt}{0.800pt}}
\put(108,96.34){\rule{1.800pt}{0.800pt}}
\multiput(108.00,94.34)(4.264,4.000){2}{\rule{0.900pt}{0.800pt}}
\put(116,100.34){\rule{1.800pt}{0.800pt}}
\multiput(116.00,98.34)(4.264,4.000){2}{\rule{0.900pt}{0.800pt}}
\put(124,104.34){\rule{1.600pt}{0.800pt}}
\multiput(124.00,102.34)(3.679,4.000){2}{\rule{0.800pt}{0.800pt}}
\put(131,108.34){\rule{1.800pt}{0.800pt}}
\multiput(131.00,106.34)(4.264,4.000){2}{\rule{0.900pt}{0.800pt}}
\put(139,111.84){\rule{1.927pt}{0.800pt}}
\multiput(139.00,110.34)(4.000,3.000){2}{\rule{0.964pt}{0.800pt}}
\put(147,115.34){\rule{1.800pt}{0.800pt}}
\multiput(147.00,113.34)(4.264,4.000){2}{\rule{0.900pt}{0.800pt}}
\put(155,119.34){\rule{1.800pt}{0.800pt}}
\multiput(155.00,117.34)(4.264,4.000){2}{\rule{0.900pt}{0.800pt}}
\put(163,123.34){\rule{1.800pt}{0.800pt}}
\multiput(163.00,121.34)(4.264,4.000){2}{\rule{0.900pt}{0.800pt}}
\put(171,127.34){\rule{1.800pt}{0.800pt}}
\multiput(171.00,125.34)(4.264,4.000){2}{\rule{0.900pt}{0.800pt}}
\put(179,131.34){\rule{1.800pt}{0.800pt}}
\multiput(179.00,129.34)(4.264,4.000){2}{\rule{0.900pt}{0.800pt}}
\put(187,135.34){\rule{1.600pt}{0.800pt}}
\multiput(187.00,133.34)(3.679,4.000){2}{\rule{0.800pt}{0.800pt}}
\put(194,138.84){\rule{1.927pt}{0.800pt}}
\multiput(194.00,137.34)(4.000,3.000){2}{\rule{0.964pt}{0.800pt}}
\put(202,142.34){\rule{1.800pt}{0.800pt}}
\multiput(202.00,140.34)(4.264,4.000){2}{\rule{0.900pt}{0.800pt}}
\put(210,146.34){\rule{1.800pt}{0.800pt}}
\multiput(210.00,144.34)(4.264,4.000){2}{\rule{0.900pt}{0.800pt}}
\put(218,150.34){\rule{1.800pt}{0.800pt}}
\multiput(218.00,148.34)(4.264,4.000){2}{\rule{0.900pt}{0.800pt}}
\put(226,154.34){\rule{1.800pt}{0.800pt}}
\multiput(226.00,152.34)(4.264,4.000){2}{\rule{0.900pt}{0.800pt}}
\put(234,158.34){\rule{1.800pt}{0.800pt}}
\multiput(234.00,156.34)(4.264,4.000){2}{\rule{0.900pt}{0.800pt}}
\put(242,162.34){\rule{1.800pt}{0.800pt}}
\multiput(242.00,160.34)(4.264,4.000){2}{\rule{0.900pt}{0.800pt}}
\put(250,166.34){\rule{1.600pt}{0.800pt}}
\multiput(250.00,164.34)(3.679,4.000){2}{\rule{0.800pt}{0.800pt}}
\put(257,170.34){\rule{1.800pt}{0.800pt}}
\multiput(257.00,168.34)(4.264,4.000){2}{\rule{0.900pt}{0.800pt}}
\put(265,174.34){\rule{1.800pt}{0.800pt}}
\multiput(265.00,172.34)(4.264,4.000){2}{\rule{0.900pt}{0.800pt}}
\put(273,178.34){\rule{1.800pt}{0.800pt}}
\multiput(273.00,176.34)(4.264,4.000){2}{\rule{0.900pt}{0.800pt}}
\put(281,182.34){\rule{1.800pt}{0.800pt}}
\multiput(281.00,180.34)(4.264,4.000){2}{\rule{0.900pt}{0.800pt}}
\put(289,186.34){\rule{1.800pt}{0.800pt}}
\multiput(289.00,184.34)(4.264,4.000){2}{\rule{0.900pt}{0.800pt}}
\put(297,190.34){\rule{1.800pt}{0.800pt}}
\multiput(297.00,188.34)(4.264,4.000){2}{\rule{0.900pt}{0.800pt}}
\put(305,194.34){\rule{1.600pt}{0.800pt}}
\multiput(305.00,192.34)(3.679,4.000){2}{\rule{0.800pt}{0.800pt}}
\put(312,198.34){\rule{1.800pt}{0.800pt}}
\multiput(312.00,196.34)(4.264,4.000){2}{\rule{0.900pt}{0.800pt}}
\put(320,202.34){\rule{1.800pt}{0.800pt}}
\multiput(320.00,200.34)(4.264,4.000){2}{\rule{0.900pt}{0.800pt}}
\put(328,206.34){\rule{1.800pt}{0.800pt}}
\multiput(328.00,204.34)(4.264,4.000){2}{\rule{0.900pt}{0.800pt}}
\multiput(336.00,211.38)(0.928,0.560){3}{\rule{1.480pt}{0.135pt}}
\multiput(336.00,208.34)(4.928,5.000){2}{\rule{0.740pt}{0.800pt}}
\put(344,215.34){\rule{1.800pt}{0.800pt}}
\multiput(344.00,213.34)(4.264,4.000){2}{\rule{0.900pt}{0.800pt}}
\put(352,219.34){\rule{1.800pt}{0.800pt}}
\multiput(352.00,217.34)(4.264,4.000){2}{\rule{0.900pt}{0.800pt}}
\put(360,223.34){\rule{1.800pt}{0.800pt}}
\multiput(360.00,221.34)(4.264,4.000){2}{\rule{0.900pt}{0.800pt}}
\multiput(368.00,228.38)(0.760,0.560){3}{\rule{1.320pt}{0.135pt}}
\multiput(368.00,225.34)(4.260,5.000){2}{\rule{0.660pt}{0.800pt}}
\put(375,232.34){\rule{1.800pt}{0.800pt}}
\multiput(375.00,230.34)(4.264,4.000){2}{\rule{0.900pt}{0.800pt}}
\multiput(383.00,237.38)(0.928,0.560){3}{\rule{1.480pt}{0.135pt}}
\multiput(383.00,234.34)(4.928,5.000){2}{\rule{0.740pt}{0.800pt}}
\put(391,241.34){\rule{1.800pt}{0.800pt}}
\multiput(391.00,239.34)(4.264,4.000){2}{\rule{0.900pt}{0.800pt}}
\multiput(399.00,246.38)(0.928,0.560){3}{\rule{1.480pt}{0.135pt}}
\multiput(399.00,243.34)(4.928,5.000){2}{\rule{0.740pt}{0.800pt}}
\multiput(407.00,251.38)(0.928,0.560){3}{\rule{1.480pt}{0.135pt}}
\multiput(407.00,248.34)(4.928,5.000){2}{\rule{0.740pt}{0.800pt}}
\multiput(415.00,256.38)(0.928,0.560){3}{\rule{1.480pt}{0.135pt}}
\multiput(415.00,253.34)(4.928,5.000){2}{\rule{0.740pt}{0.800pt}}
\put(423,260.34){\rule{1.600pt}{0.800pt}}
\multiput(423.00,258.34)(3.679,4.000){2}{\rule{0.800pt}{0.800pt}}
\multiput(430.00,265.38)(0.928,0.560){3}{\rule{1.480pt}{0.135pt}}
\multiput(430.00,262.34)(4.928,5.000){2}{\rule{0.740pt}{0.800pt}}
\multiput(438.00,270.38)(0.928,0.560){3}{\rule{1.480pt}{0.135pt}}
\multiput(438.00,267.34)(4.928,5.000){2}{\rule{0.740pt}{0.800pt}}
\multiput(446.00,275.38)(0.928,0.560){3}{\rule{1.480pt}{0.135pt}}
\multiput(446.00,272.34)(4.928,5.000){2}{\rule{0.740pt}{0.800pt}}
\multiput(454.00,280.38)(0.928,0.560){3}{\rule{1.480pt}{0.135pt}}
\multiput(454.00,277.34)(4.928,5.000){2}{\rule{0.740pt}{0.800pt}}
\multiput(462.00,285.39)(0.685,0.536){5}{\rule{1.267pt}{0.129pt}}
\multiput(462.00,282.34)(5.371,6.000){2}{\rule{0.633pt}{0.800pt}}
\multiput(470.00,291.38)(0.928,0.560){3}{\rule{1.480pt}{0.135pt}}
\multiput(470.00,288.34)(4.928,5.000){2}{\rule{0.740pt}{0.800pt}}
\multiput(478.00,296.38)(0.928,0.560){3}{\rule{1.480pt}{0.135pt}}
\multiput(478.00,293.34)(4.928,5.000){2}{\rule{0.740pt}{0.800pt}}
\multiput(486.00,301.39)(0.574,0.536){5}{\rule{1.133pt}{0.129pt}}
\multiput(486.00,298.34)(4.648,6.000){2}{\rule{0.567pt}{0.800pt}}
\multiput(493.00,307.38)(0.928,0.560){3}{\rule{1.480pt}{0.135pt}}
\multiput(493.00,304.34)(4.928,5.000){2}{\rule{0.740pt}{0.800pt}}
\multiput(501.00,312.39)(0.685,0.536){5}{\rule{1.267pt}{0.129pt}}
\multiput(501.00,309.34)(5.371,6.000){2}{\rule{0.633pt}{0.800pt}}
\multiput(509.00,318.39)(0.685,0.536){5}{\rule{1.267pt}{0.129pt}}
\multiput(509.00,315.34)(5.371,6.000){2}{\rule{0.633pt}{0.800pt}}
\multiput(517.00,324.39)(0.685,0.536){5}{\rule{1.267pt}{0.129pt}}
\multiput(517.00,321.34)(5.371,6.000){2}{\rule{0.633pt}{0.800pt}}
\multiput(525.00,330.39)(0.685,0.536){5}{\rule{1.267pt}{0.129pt}}
\multiput(525.00,327.34)(5.371,6.000){2}{\rule{0.633pt}{0.800pt}}
\multiput(533.00,336.39)(0.685,0.536){5}{\rule{1.267pt}{0.129pt}}
\multiput(533.00,333.34)(5.371,6.000){2}{\rule{0.633pt}{0.800pt}}
\multiput(541.00,342.39)(0.685,0.536){5}{\rule{1.267pt}{0.129pt}}
\multiput(541.00,339.34)(5.371,6.000){2}{\rule{0.633pt}{0.800pt}}
\multiput(549.00,348.39)(0.574,0.536){5}{\rule{1.133pt}{0.129pt}}
\multiput(549.00,345.34)(4.648,6.000){2}{\rule{0.567pt}{0.800pt}}
\multiput(556.00,354.40)(0.562,0.526){7}{\rule{1.114pt}{0.127pt}}
\multiput(556.00,351.34)(5.687,7.000){2}{\rule{0.557pt}{0.800pt}}
\multiput(564.00,361.40)(0.562,0.526){7}{\rule{1.114pt}{0.127pt}}
\multiput(564.00,358.34)(5.687,7.000){2}{\rule{0.557pt}{0.800pt}}
\multiput(572.00,368.39)(0.685,0.536){5}{\rule{1.267pt}{0.129pt}}
\multiput(572.00,365.34)(5.371,6.000){2}{\rule{0.633pt}{0.800pt}}
\multiput(580.00,374.40)(0.562,0.526){7}{\rule{1.114pt}{0.127pt}}
\multiput(580.00,371.34)(5.687,7.000){2}{\rule{0.557pt}{0.800pt}}
\multiput(588.00,381.40)(0.562,0.526){7}{\rule{1.114pt}{0.127pt}}
\multiput(588.00,378.34)(5.687,7.000){2}{\rule{0.557pt}{0.800pt}}
\multiput(596.00,388.40)(0.562,0.526){7}{\rule{1.114pt}{0.127pt}}
\multiput(596.00,385.34)(5.687,7.000){2}{\rule{0.557pt}{0.800pt}}
\multiput(605.40,394.00)(0.526,0.562){7}{\rule{0.127pt}{1.114pt}}
\multiput(602.34,394.00)(7.000,5.687){2}{\rule{0.800pt}{0.557pt}}
\multiput(611.00,403.40)(0.562,0.526){7}{\rule{1.114pt}{0.127pt}}
\multiput(611.00,400.34)(5.687,7.000){2}{\rule{0.557pt}{0.800pt}}
\multiput(619.00,410.40)(0.481,0.520){9}{\rule{1.000pt}{0.125pt}}
\multiput(619.00,407.34)(5.924,8.000){2}{\rule{0.500pt}{0.800pt}}
\multiput(627.00,418.40)(0.562,0.526){7}{\rule{1.114pt}{0.127pt}}
\multiput(627.00,415.34)(5.687,7.000){2}{\rule{0.557pt}{0.800pt}}
\multiput(635.00,425.40)(0.481,0.520){9}{\rule{1.000pt}{0.125pt}}
\multiput(635.00,422.34)(5.924,8.000){2}{\rule{0.500pt}{0.800pt}}
\multiput(643.00,433.40)(0.481,0.520){9}{\rule{1.000pt}{0.125pt}}
\multiput(643.00,430.34)(5.924,8.000){2}{\rule{0.500pt}{0.800pt}}
\multiput(652.40,440.00)(0.520,0.554){9}{\rule{0.125pt}{1.100pt}}
\multiput(649.34,440.00)(8.000,6.717){2}{\rule{0.800pt}{0.550pt}}
\multiput(659.00,450.40)(0.481,0.520){9}{\rule{1.000pt}{0.125pt}}
\multiput(659.00,447.34)(5.924,8.000){2}{\rule{0.500pt}{0.800pt}}
\multiput(668.40,457.00)(0.526,0.650){7}{\rule{0.127pt}{1.229pt}}
\multiput(665.34,457.00)(7.000,6.450){2}{\rule{0.800pt}{0.614pt}}
\multiput(674.00,467.40)(0.481,0.520){9}{\rule{1.000pt}{0.125pt}}
\multiput(674.00,464.34)(5.924,8.000){2}{\rule{0.500pt}{0.800pt}}
\multiput(683.40,474.00)(0.520,0.554){9}{\rule{0.125pt}{1.100pt}}
\multiput(680.34,474.00)(8.000,6.717){2}{\rule{0.800pt}{0.550pt}}
\multiput(691.40,483.00)(0.520,0.554){9}{\rule{0.125pt}{1.100pt}}
\multiput(688.34,483.00)(8.000,6.717){2}{\rule{0.800pt}{0.550pt}}
\multiput(699.40,492.00)(0.520,0.627){9}{\rule{0.125pt}{1.200pt}}
\multiput(696.34,492.00)(8.000,7.509){2}{\rule{0.800pt}{0.600pt}}
\multiput(707.40,502.00)(0.520,0.554){9}{\rule{0.125pt}{1.100pt}}
\multiput(704.34,502.00)(8.000,6.717){2}{\rule{0.800pt}{0.550pt}}
\multiput(715.40,511.00)(0.520,0.627){9}{\rule{0.125pt}{1.200pt}}
\multiput(712.34,511.00)(8.000,7.509){2}{\rule{0.800pt}{0.600pt}}
\multiput(723.40,521.00)(0.526,0.738){7}{\rule{0.127pt}{1.343pt}}
\multiput(720.34,521.00)(7.000,7.213){2}{\rule{0.800pt}{0.671pt}}
\multiput(730.40,531.00)(0.520,0.627){9}{\rule{0.125pt}{1.200pt}}
\multiput(727.34,531.00)(8.000,7.509){2}{\rule{0.800pt}{0.600pt}}
\multiput(738.40,541.00)(0.520,0.700){9}{\rule{0.125pt}{1.300pt}}
\multiput(735.34,541.00)(8.000,8.302){2}{\rule{0.800pt}{0.650pt}}
\put(745,551.34){\rule{0.482pt}{0.800pt}}
\multiput(745.00,550.34)(1.000,2.000){2}{\rule{0.241pt}{0.800pt}}
\sbox{\plotpoint}{\rule[-0.200pt]{0.400pt}{0.400pt}}%
\put(506,82){\usebox{\plotpoint}}
\put(506.0,82.0){\rule[-0.200pt]{0.400pt}{113.705pt}}
\put(701,82){\usebox{\plotpoint}}
\put(701.0,82.0){\rule[-0.200pt]{0.400pt}{113.705pt}}
\sbox{\plotpoint}{\rule[-0.400pt]{0.800pt}{0.800pt}}%
\put(506,172){\usebox{\plotpoint}}
\multiput(506.00,173.39)(1.913,0.536){5}{\rule{2.733pt}{0.129pt}}
\multiput(506.00,170.34)(13.327,6.000){2}{\rule{1.367pt}{0.800pt}}
\multiput(525.00,179.39)(2.137,0.536){5}{\rule{3.000pt}{0.129pt}}
\multiput(525.00,176.34)(14.773,6.000){2}{\rule{1.500pt}{0.800pt}}
\multiput(546.00,185.39)(2.137,0.536){5}{\rule{3.000pt}{0.129pt}}
\multiput(546.00,182.34)(14.773,6.000){2}{\rule{1.500pt}{0.800pt}}
\multiput(567.00,191.39)(2.248,0.536){5}{\rule{3.133pt}{0.129pt}}
\multiput(567.00,188.34)(15.497,6.000){2}{\rule{1.567pt}{0.800pt}}
\multiput(589.00,197.38)(3.111,0.560){3}{\rule{3.560pt}{0.135pt}}
\multiput(589.00,194.34)(13.611,5.000){2}{\rule{1.780pt}{0.800pt}}
\put(610,201.34){\rule{4.400pt}{0.800pt}}
\multiput(610.00,199.34)(11.868,4.000){2}{\rule{2.200pt}{0.800pt}}
\put(631,204.84){\rule{5.059pt}{0.800pt}}
\multiput(631.00,203.34)(10.500,3.000){2}{\rule{2.529pt}{0.800pt}}
\put(652,206.84){\rule{5.300pt}{0.800pt}}
\multiput(652.00,206.34)(11.000,1.000){2}{\rule{2.650pt}{0.800pt}}
\put(674,205.84){\rule{6.504pt}{0.800pt}}
\multiput(674.00,207.34)(13.500,-3.000){2}{\rule{3.252pt}{0.800pt}}
\end{picture}
{\small FIG.\ 1: Reduced $h_{KS}$ values for hard sphere fluids,
labelled as in the text. 
Here $m$ is the mass of the particles, $n$ their number density,
$\sigma$ their radius, $N$ their number and $K$ 
their kinetic energy. The range of $n\sigma^3$ used by DAV 
is indicated by the vertical lines.} 
\end{figure}
\noindent

Paper [1] was inspired by an earlier one of Dzugutov [3]
concerning a universal
connection between the self diffusion coefficent $D$ and $S_{ex}$  
of a number 
of dense (including hard sphere) fluids, of the form:
$D^*\equiv D/\Gamma \sigma^2 = 0.049 e^{S_{ex}}$ (2), where $\sigma$ is 
an effective hard sphere diameter. 
The range of density considered in [2] is $0.4 \le n \sigma^3 
\le 0.9$, i.e.\
restricted to high fluid densities, but as shown in Fig.2 the agreement
of (2) and the numerical data from Alder, Gass and Wainwright (AGW) [4] 
and Erpenbeck and Wood (EW) [5] for hard spheres is not good. 
The problem is shown in a wider range in
Fig.\ 3 where (2), transformed to $D/D_B$, is compared 
with the numerical data of [4,5].
Here $D_B$ is the Boltzmann value for the
self diffusion coefficient.
Note that $D/D_B$ does not monotonically decrease with density,
which is unphysical.
Nevertheless, a connection between dynamical ($h_{KS}$, $D$)
and thermodynamical ($S_{ex}$) quantities remains a 
challenge.

The authors would like to thank I.M.\ de Schepper for very elpful 
correspondence.

\vskip 5pt\noindent
\begin{figure}
\setlength{\unitlength}{0.240900pt}
\ifx\plotpoint\undefined\newsavebox{\plotpoint}\fi
\sbox{\plotpoint}{\rule[-0.200pt]{0.400pt}{0.400pt}}%
\begin{picture}(900,594)(0,0)
\font\gnuplot=cmr10 at 10pt
\gnuplot
\sbox{\plotpoint}{\rule[-0.200pt]{0.400pt}{0.400pt}}%
\put(140.0,82.0){\rule[-0.200pt]{4.818pt}{0.400pt}}
\put(120,82){\makebox(0,0)[r]{0.00}}
\put(859.0,82.0){\rule[-0.200pt]{4.818pt}{0.400pt}}
\put(140.0,292.0){\rule[-0.200pt]{4.818pt}{0.400pt}}
\put(120,292){\makebox(0,0)[r]{0.02}}
\put(859.0,292.0){\rule[-0.200pt]{4.818pt}{0.400pt}}
\put(140.0,502.0){\rule[-0.200pt]{4.818pt}{0.400pt}}
\put(120,502){\makebox(0,0)[r]{0.04}}
\put(859.0,502.0){\rule[-0.200pt]{4.818pt}{0.400pt}}
\put(161.0,82.0){\rule[-0.200pt]{0.400pt}{4.818pt}}
\put(161,41){\makebox(0,0){1.5}}
\put(161.0,534.0){\rule[-0.200pt]{0.400pt}{4.818pt}}
\put(366.0,82.0){\rule[-0.200pt]{0.400pt}{4.818pt}}
\put(366,41){\makebox(0,0){2.5}}
\put(366.0,534.0){\rule[-0.200pt]{0.400pt}{4.818pt}}
\put(571.0,82.0){\rule[-0.200pt]{0.400pt}{4.818pt}}
\put(571,41){\makebox(0,0){3.5}}
\put(571.0,534.0){\rule[-0.200pt]{0.400pt}{4.818pt}}
\put(776.0,82.0){\rule[-0.200pt]{0.400pt}{4.818pt}}
\put(776,41){\makebox(0,0){4.5}}
\put(776.0,534.0){\rule[-0.200pt]{0.400pt}{4.818pt}}
\put(140.0,82.0){\rule[-0.200pt]{178.025pt}{0.400pt}}
\put(879.0,82.0){\rule[-0.200pt]{0.400pt}{113.705pt}}
\put(140.0,554.0){\rule[-0.200pt]{178.025pt}{0.400pt}}
\put(778,498){\makebox(0,0){$D^*$}}
\put(469,21){\makebox(0,0){$- S_{ex}$}}
\put(161,218){\makebox(0,0)[l]{Dz.}}
\put(243,344){\makebox(0,0)[l]{AGW,EW}}
\put(140.0,82.0){\rule[-0.200pt]{0.400pt}{113.705pt}}
\put(140,209){\usebox{\plotpoint}}
\multiput(140.00,207.93)(0.710,-0.477){7}{\rule{0.660pt}{0.115pt}}
\multiput(140.00,208.17)(5.630,-5.000){2}{\rule{0.330pt}{0.400pt}}
\multiput(147.00,202.94)(1.066,-0.468){5}{\rule{0.900pt}{0.113pt}}
\multiput(147.00,203.17)(6.132,-4.000){2}{\rule{0.450pt}{0.400pt}}
\multiput(155.00,198.94)(0.920,-0.468){5}{\rule{0.800pt}{0.113pt}}
\multiput(155.00,199.17)(5.340,-4.000){2}{\rule{0.400pt}{0.400pt}}
\multiput(162.00,194.94)(1.066,-0.468){5}{\rule{0.900pt}{0.113pt}}
\multiput(162.00,195.17)(6.132,-4.000){2}{\rule{0.450pt}{0.400pt}}
\multiput(170.00,190.94)(0.920,-0.468){5}{\rule{0.800pt}{0.113pt}}
\multiput(170.00,191.17)(5.340,-4.000){2}{\rule{0.400pt}{0.400pt}}
\multiput(177.00,186.94)(1.066,-0.468){5}{\rule{0.900pt}{0.113pt}}
\multiput(177.00,187.17)(6.132,-4.000){2}{\rule{0.450pt}{0.400pt}}
\multiput(185.00,182.94)(0.920,-0.468){5}{\rule{0.800pt}{0.113pt}}
\multiput(185.00,183.17)(5.340,-4.000){2}{\rule{0.400pt}{0.400pt}}
\multiput(192.00,178.95)(1.579,-0.447){3}{\rule{1.167pt}{0.108pt}}
\multiput(192.00,179.17)(5.579,-3.000){2}{\rule{0.583pt}{0.400pt}}
\multiput(200.00,175.94)(0.920,-0.468){5}{\rule{0.800pt}{0.113pt}}
\multiput(200.00,176.17)(5.340,-4.000){2}{\rule{0.400pt}{0.400pt}}
\multiput(207.00,171.95)(1.579,-0.447){3}{\rule{1.167pt}{0.108pt}}
\multiput(207.00,172.17)(5.579,-3.000){2}{\rule{0.583pt}{0.400pt}}
\multiput(215.00,168.95)(1.355,-0.447){3}{\rule{1.033pt}{0.108pt}}
\multiput(215.00,169.17)(4.855,-3.000){2}{\rule{0.517pt}{0.400pt}}
\multiput(222.00,165.95)(1.579,-0.447){3}{\rule{1.167pt}{0.108pt}}
\multiput(222.00,166.17)(5.579,-3.000){2}{\rule{0.583pt}{0.400pt}}
\multiput(230.00,162.95)(1.355,-0.447){3}{\rule{1.033pt}{0.108pt}}
\multiput(230.00,163.17)(4.855,-3.000){2}{\rule{0.517pt}{0.400pt}}
\multiput(237.00,159.95)(1.579,-0.447){3}{\rule{1.167pt}{0.108pt}}
\multiput(237.00,160.17)(5.579,-3.000){2}{\rule{0.583pt}{0.400pt}}
\multiput(245.00,156.95)(1.355,-0.447){3}{\rule{1.033pt}{0.108pt}}
\multiput(245.00,157.17)(4.855,-3.000){2}{\rule{0.517pt}{0.400pt}}
\put(252,153.17){\rule{1.500pt}{0.400pt}}
\multiput(252.00,154.17)(3.887,-2.000){2}{\rule{0.750pt}{0.400pt}}
\multiput(259.00,151.95)(1.579,-0.447){3}{\rule{1.167pt}{0.108pt}}
\multiput(259.00,152.17)(5.579,-3.000){2}{\rule{0.583pt}{0.400pt}}
\put(267,148.17){\rule{1.500pt}{0.400pt}}
\multiput(267.00,149.17)(3.887,-2.000){2}{\rule{0.750pt}{0.400pt}}
\put(274,146.17){\rule{1.700pt}{0.400pt}}
\multiput(274.00,147.17)(4.472,-2.000){2}{\rule{0.850pt}{0.400pt}}
\multiput(282.00,144.95)(1.355,-0.447){3}{\rule{1.033pt}{0.108pt}}
\multiput(282.00,145.17)(4.855,-3.000){2}{\rule{0.517pt}{0.400pt}}
\put(289,141.17){\rule{1.700pt}{0.400pt}}
\multiput(289.00,142.17)(4.472,-2.000){2}{\rule{0.850pt}{0.400pt}}
\put(297,139.17){\rule{1.500pt}{0.400pt}}
\multiput(297.00,140.17)(3.887,-2.000){2}{\rule{0.750pt}{0.400pt}}
\put(304,137.17){\rule{1.700pt}{0.400pt}}
\multiput(304.00,138.17)(4.472,-2.000){2}{\rule{0.850pt}{0.400pt}}
\put(312,135.17){\rule{1.500pt}{0.400pt}}
\multiput(312.00,136.17)(3.887,-2.000){2}{\rule{0.750pt}{0.400pt}}
\put(319,133.17){\rule{1.700pt}{0.400pt}}
\multiput(319.00,134.17)(4.472,-2.000){2}{\rule{0.850pt}{0.400pt}}
\put(327,131.17){\rule{1.500pt}{0.400pt}}
\multiput(327.00,132.17)(3.887,-2.000){2}{\rule{0.750pt}{0.400pt}}
\put(334,129.17){\rule{1.700pt}{0.400pt}}
\multiput(334.00,130.17)(4.472,-2.000){2}{\rule{0.850pt}{0.400pt}}
\put(342,127.67){\rule{1.686pt}{0.400pt}}
\multiput(342.00,128.17)(3.500,-1.000){2}{\rule{0.843pt}{0.400pt}}
\put(349,126.17){\rule{1.500pt}{0.400pt}}
\multiput(349.00,127.17)(3.887,-2.000){2}{\rule{0.750pt}{0.400pt}}
\put(356,124.67){\rule{1.927pt}{0.400pt}}
\multiput(356.00,125.17)(4.000,-1.000){2}{\rule{0.964pt}{0.400pt}}
\put(364,123.17){\rule{1.500pt}{0.400pt}}
\multiput(364.00,124.17)(3.887,-2.000){2}{\rule{0.750pt}{0.400pt}}
\put(371,121.67){\rule{1.927pt}{0.400pt}}
\multiput(371.00,122.17)(4.000,-1.000){2}{\rule{0.964pt}{0.400pt}}
\put(379,120.17){\rule{1.500pt}{0.400pt}}
\multiput(379.00,121.17)(3.887,-2.000){2}{\rule{0.750pt}{0.400pt}}
\put(386,118.67){\rule{1.927pt}{0.400pt}}
\multiput(386.00,119.17)(4.000,-1.000){2}{\rule{0.964pt}{0.400pt}}
\put(394,117.17){\rule{1.500pt}{0.400pt}}
\multiput(394.00,118.17)(3.887,-2.000){2}{\rule{0.750pt}{0.400pt}}
\put(401,115.67){\rule{1.927pt}{0.400pt}}
\multiput(401.00,116.17)(4.000,-1.000){2}{\rule{0.964pt}{0.400pt}}
\put(409,114.67){\rule{1.686pt}{0.400pt}}
\multiput(409.00,115.17)(3.500,-1.000){2}{\rule{0.843pt}{0.400pt}}
\put(416,113.67){\rule{1.927pt}{0.400pt}}
\multiput(416.00,114.17)(4.000,-1.000){2}{\rule{0.964pt}{0.400pt}}
\put(424,112.67){\rule{1.686pt}{0.400pt}}
\multiput(424.00,113.17)(3.500,-1.000){2}{\rule{0.843pt}{0.400pt}}
\put(431,111.67){\rule{1.927pt}{0.400pt}}
\multiput(431.00,112.17)(4.000,-1.000){2}{\rule{0.964pt}{0.400pt}}
\put(439,110.67){\rule{1.686pt}{0.400pt}}
\multiput(439.00,111.17)(3.500,-1.000){2}{\rule{0.843pt}{0.400pt}}
\put(446,109.67){\rule{1.927pt}{0.400pt}}
\multiput(446.00,110.17)(4.000,-1.000){2}{\rule{0.964pt}{0.400pt}}
\put(454,108.67){\rule{1.686pt}{0.400pt}}
\multiput(454.00,109.17)(3.500,-1.000){2}{\rule{0.843pt}{0.400pt}}
\put(461,107.67){\rule{1.686pt}{0.400pt}}
\multiput(461.00,108.17)(3.500,-1.000){2}{\rule{0.843pt}{0.400pt}}
\put(468,106.67){\rule{1.927pt}{0.400pt}}
\multiput(468.00,107.17)(4.000,-1.000){2}{\rule{0.964pt}{0.400pt}}
\put(476,105.67){\rule{1.686pt}{0.400pt}}
\multiput(476.00,106.17)(3.500,-1.000){2}{\rule{0.843pt}{0.400pt}}
\put(483,104.67){\rule{1.927pt}{0.400pt}}
\multiput(483.00,105.17)(4.000,-1.000){2}{\rule{0.964pt}{0.400pt}}
\put(491,103.67){\rule{1.686pt}{0.400pt}}
\multiput(491.00,104.17)(3.500,-1.000){2}{\rule{0.843pt}{0.400pt}}
\put(498,102.67){\rule{1.927pt}{0.400pt}}
\multiput(498.00,103.17)(4.000,-1.000){2}{\rule{0.964pt}{0.400pt}}
\put(513,101.67){\rule{1.927pt}{0.400pt}}
\multiput(513.00,102.17)(4.000,-1.000){2}{\rule{0.964pt}{0.400pt}}
\put(521,100.67){\rule{1.686pt}{0.400pt}}
\multiput(521.00,101.17)(3.500,-1.000){2}{\rule{0.843pt}{0.400pt}}
\put(528,99.67){\rule{1.927pt}{0.400pt}}
\multiput(528.00,100.17)(4.000,-1.000){2}{\rule{0.964pt}{0.400pt}}
\put(506.0,103.0){\rule[-0.200pt]{1.686pt}{0.400pt}}
\put(543,98.67){\rule{1.927pt}{0.400pt}}
\multiput(543.00,99.17)(4.000,-1.000){2}{\rule{0.964pt}{0.400pt}}
\put(536.0,100.0){\rule[-0.200pt]{1.686pt}{0.400pt}}
\put(558,97.67){\rule{1.686pt}{0.400pt}}
\multiput(558.00,98.17)(3.500,-1.000){2}{\rule{0.843pt}{0.400pt}}
\put(565,96.67){\rule{1.927pt}{0.400pt}}
\multiput(565.00,97.17)(4.000,-1.000){2}{\rule{0.964pt}{0.400pt}}
\put(551.0,99.0){\rule[-0.200pt]{1.686pt}{0.400pt}}
\put(580,95.67){\rule{1.927pt}{0.400pt}}
\multiput(580.00,96.17)(4.000,-1.000){2}{\rule{0.964pt}{0.400pt}}
\put(573.0,97.0){\rule[-0.200pt]{1.686pt}{0.400pt}}
\put(595,94.67){\rule{1.927pt}{0.400pt}}
\multiput(595.00,95.17)(4.000,-1.000){2}{\rule{0.964pt}{0.400pt}}
\put(588.0,96.0){\rule[-0.200pt]{1.686pt}{0.400pt}}
\put(610,93.67){\rule{1.927pt}{0.400pt}}
\multiput(610.00,94.17)(4.000,-1.000){2}{\rule{0.964pt}{0.400pt}}
\put(603.0,95.0){\rule[-0.200pt]{1.686pt}{0.400pt}}
\put(625,92.67){\rule{1.927pt}{0.400pt}}
\multiput(625.00,93.17)(4.000,-1.000){2}{\rule{0.964pt}{0.400pt}}
\put(618.0,94.0){\rule[-0.200pt]{1.686pt}{0.400pt}}
\put(648,91.67){\rule{1.686pt}{0.400pt}}
\multiput(648.00,92.17)(3.500,-1.000){2}{\rule{0.843pt}{0.400pt}}
\put(633.0,93.0){\rule[-0.200pt]{3.613pt}{0.400pt}}
\put(670,90.67){\rule{1.686pt}{0.400pt}}
\multiput(670.00,91.17)(3.500,-1.000){2}{\rule{0.843pt}{0.400pt}}
\put(655.0,92.0){\rule[-0.200pt]{3.613pt}{0.400pt}}
\put(692,89.67){\rule{1.927pt}{0.400pt}}
\multiput(692.00,90.17)(4.000,-1.000){2}{\rule{0.964pt}{0.400pt}}
\put(677.0,91.0){\rule[-0.200pt]{3.613pt}{0.400pt}}
\put(715,88.67){\rule{1.686pt}{0.400pt}}
\multiput(715.00,89.17)(3.500,-1.000){2}{\rule{0.843pt}{0.400pt}}
\put(700.0,90.0){\rule[-0.200pt]{3.613pt}{0.400pt}}
\put(745,87.67){\rule{1.686pt}{0.400pt}}
\multiput(745.00,88.17)(3.500,-1.000){2}{\rule{0.843pt}{0.400pt}}
\put(722.0,89.0){\rule[-0.200pt]{5.541pt}{0.400pt}}
\put(782,86.67){\rule{1.686pt}{0.400pt}}
\multiput(782.00,87.17)(3.500,-1.000){2}{\rule{0.843pt}{0.400pt}}
\put(752.0,88.0){\rule[-0.200pt]{7.227pt}{0.400pt}}
\put(819,85.67){\rule{1.927pt}{0.400pt}}
\multiput(819.00,86.17)(4.000,-1.000){2}{\rule{0.964pt}{0.400pt}}
\put(789.0,87.0){\rule[-0.200pt]{7.227pt}{0.400pt}}
\put(872,84.67){\rule{1.686pt}{0.400pt}}
\multiput(872.00,85.17)(3.500,-1.000){2}{\rule{0.843pt}{0.400pt}}
\put(827.0,86.0){\rule[-0.200pt]{10.840pt}{0.400pt}}
\put(144,529){\raisebox{-.8pt}{\makebox(0,0){$\Diamond$}}}
\put(406,157){\raisebox{-.8pt}{\makebox(0,0){$\Diamond$}}}
\put(526,121){\raisebox{-.8pt}{\makebox(0,0){$\Diamond$}}}
\put(610,106){\raisebox{-.8pt}{\makebox(0,0){$\Diamond$}}}
\put(713,96){\raisebox{-.8pt}{\makebox(0,0){$\Diamond$}}}
\put(850,88){\raisebox{-.8pt}{\makebox(0,0){$\Diamond$}}}
\sbox{\plotpoint}{\rule[-0.400pt]{0.800pt}{0.800pt}}%
\put(140,543){\usebox{\plotpoint}}
\multiput(141.40,530.78)(0.526,-1.964){7}{\rule{0.127pt}{2.943pt}}
\multiput(138.34,536.89)(7.000,-17.892){2}{\rule{0.800pt}{1.471pt}}
\multiput(148.40,508.62)(0.520,-1.578){9}{\rule{0.125pt}{2.500pt}}
\multiput(145.34,513.81)(8.000,-17.811){2}{\rule{0.800pt}{1.250pt}}
\multiput(156.40,484.73)(0.526,-1.789){7}{\rule{0.127pt}{2.714pt}}
\multiput(153.34,490.37)(7.000,-16.366){2}{\rule{0.800pt}{1.357pt}}
\multiput(163.40,464.87)(0.520,-1.358){9}{\rule{0.125pt}{2.200pt}}
\multiput(160.34,469.43)(8.000,-15.434){2}{\rule{0.800pt}{1.100pt}}
\multiput(171.40,443.68)(0.526,-1.614){7}{\rule{0.127pt}{2.486pt}}
\multiput(168.34,448.84)(7.000,-14.841){2}{\rule{0.800pt}{1.243pt}}
\multiput(178.40,425.70)(0.520,-1.212){9}{\rule{0.125pt}{2.000pt}}
\multiput(175.34,429.85)(8.000,-13.849){2}{\rule{0.800pt}{1.000pt}}
\multiput(186.40,406.63)(0.526,-1.438){7}{\rule{0.127pt}{2.257pt}}
\multiput(183.34,411.32)(7.000,-13.315){2}{\rule{0.800pt}{1.129pt}}
\multiput(193.40,390.53)(0.520,-1.066){9}{\rule{0.125pt}{1.800pt}}
\multiput(190.34,394.26)(8.000,-12.264){2}{\rule{0.800pt}{0.900pt}}
\multiput(201.40,373.58)(0.526,-1.263){7}{\rule{0.127pt}{2.029pt}}
\multiput(198.34,377.79)(7.000,-11.790){2}{\rule{0.800pt}{1.014pt}}
\multiput(208.40,358.94)(0.520,-0.993){9}{\rule{0.125pt}{1.700pt}}
\multiput(205.34,362.47)(8.000,-11.472){2}{\rule{0.800pt}{0.850pt}}
\multiput(216.40,343.53)(0.526,-1.088){7}{\rule{0.127pt}{1.800pt}}
\multiput(213.34,347.26)(7.000,-10.264){2}{\rule{0.800pt}{0.900pt}}
\multiput(223.40,330.77)(0.520,-0.847){9}{\rule{0.125pt}{1.500pt}}
\multiput(220.34,333.89)(8.000,-9.887){2}{\rule{0.800pt}{0.750pt}}
\multiput(231.40,317.00)(0.526,-1.000){7}{\rule{0.127pt}{1.686pt}}
\multiput(228.34,320.50)(7.000,-9.501){2}{\rule{0.800pt}{0.843pt}}
\multiput(238.40,305.19)(0.520,-0.774){9}{\rule{0.125pt}{1.400pt}}
\multiput(235.34,308.09)(8.000,-9.094){2}{\rule{0.800pt}{0.700pt}}
\multiput(246.40,292.95)(0.526,-0.825){7}{\rule{0.127pt}{1.457pt}}
\multiput(243.34,295.98)(7.000,-7.976){2}{\rule{0.800pt}{0.729pt}}
\multiput(253.40,281.95)(0.526,-0.825){7}{\rule{0.127pt}{1.457pt}}
\multiput(250.34,284.98)(7.000,-7.976){2}{\rule{0.800pt}{0.729pt}}
\multiput(260.40,272.02)(0.520,-0.627){9}{\rule{0.125pt}{1.200pt}}
\multiput(257.34,274.51)(8.000,-7.509){2}{\rule{0.800pt}{0.600pt}}
\multiput(268.40,261.90)(0.526,-0.650){7}{\rule{0.127pt}{1.229pt}}
\multiput(265.34,264.45)(7.000,-6.450){2}{\rule{0.800pt}{0.614pt}}
\multiput(275.40,253.43)(0.520,-0.554){9}{\rule{0.125pt}{1.100pt}}
\multiput(272.34,255.72)(8.000,-6.717){2}{\rule{0.800pt}{0.550pt}}
\multiput(283.40,243.90)(0.526,-0.650){7}{\rule{0.127pt}{1.229pt}}
\multiput(280.34,246.45)(7.000,-6.450){2}{\rule{0.800pt}{0.614pt}}
\multiput(289.00,238.08)(0.481,-0.520){9}{\rule{1.000pt}{0.125pt}}
\multiput(289.00,238.34)(5.924,-8.000){2}{\rule{0.500pt}{0.800pt}}
\multiput(297.00,230.08)(0.475,-0.526){7}{\rule{1.000pt}{0.127pt}}
\multiput(297.00,230.34)(4.924,-7.000){2}{\rule{0.500pt}{0.800pt}}
\multiput(304.00,223.08)(0.562,-0.526){7}{\rule{1.114pt}{0.127pt}}
\multiput(304.00,223.34)(5.687,-7.000){2}{\rule{0.557pt}{0.800pt}}
\multiput(312.00,216.08)(0.475,-0.526){7}{\rule{1.000pt}{0.127pt}}
\multiput(312.00,216.34)(4.924,-7.000){2}{\rule{0.500pt}{0.800pt}}
\multiput(319.00,209.07)(0.685,-0.536){5}{\rule{1.267pt}{0.129pt}}
\multiput(319.00,209.34)(5.371,-6.000){2}{\rule{0.633pt}{0.800pt}}
\multiput(327.00,203.07)(0.574,-0.536){5}{\rule{1.133pt}{0.129pt}}
\multiput(327.00,203.34)(4.648,-6.000){2}{\rule{0.567pt}{0.800pt}}
\multiput(334.00,197.07)(0.685,-0.536){5}{\rule{1.267pt}{0.129pt}}
\multiput(334.00,197.34)(5.371,-6.000){2}{\rule{0.633pt}{0.800pt}}
\multiput(342.00,191.06)(0.760,-0.560){3}{\rule{1.320pt}{0.135pt}}
\multiput(342.00,191.34)(4.260,-5.000){2}{\rule{0.660pt}{0.800pt}}
\multiput(349.00,186.06)(0.760,-0.560){3}{\rule{1.320pt}{0.135pt}}
\multiput(349.00,186.34)(4.260,-5.000){2}{\rule{0.660pt}{0.800pt}}
\put(356,179.34){\rule{1.800pt}{0.800pt}}
\multiput(356.00,181.34)(4.264,-4.000){2}{\rule{0.900pt}{0.800pt}}
\multiput(364.00,177.06)(0.760,-0.560){3}{\rule{1.320pt}{0.135pt}}
\multiput(364.00,177.34)(4.260,-5.000){2}{\rule{0.660pt}{0.800pt}}
\put(371,170.34){\rule{1.800pt}{0.800pt}}
\multiput(371.00,172.34)(4.264,-4.000){2}{\rule{0.900pt}{0.800pt}}
\put(379,166.34){\rule{1.600pt}{0.800pt}}
\multiput(379.00,168.34)(3.679,-4.000){2}{\rule{0.800pt}{0.800pt}}
\put(386,162.34){\rule{1.800pt}{0.800pt}}
\multiput(386.00,164.34)(4.264,-4.000){2}{\rule{0.900pt}{0.800pt}}
\put(394,158.84){\rule{1.686pt}{0.800pt}}
\multiput(394.00,160.34)(3.500,-3.000){2}{\rule{0.843pt}{0.800pt}}
\put(401,155.84){\rule{1.927pt}{0.800pt}}
\multiput(401.00,157.34)(4.000,-3.000){2}{\rule{0.964pt}{0.800pt}}
\put(409,152.34){\rule{1.600pt}{0.800pt}}
\multiput(409.00,154.34)(3.679,-4.000){2}{\rule{0.800pt}{0.800pt}}
\put(416,148.84){\rule{1.927pt}{0.800pt}}
\multiput(416.00,150.34)(4.000,-3.000){2}{\rule{0.964pt}{0.800pt}}
\put(424,146.34){\rule{1.686pt}{0.800pt}}
\multiput(424.00,147.34)(3.500,-2.000){2}{\rule{0.843pt}{0.800pt}}
\put(431,143.84){\rule{1.927pt}{0.800pt}}
\multiput(431.00,145.34)(4.000,-3.000){2}{\rule{0.964pt}{0.800pt}}
\put(439,140.84){\rule{1.686pt}{0.800pt}}
\multiput(439.00,142.34)(3.500,-3.000){2}{\rule{0.843pt}{0.800pt}}
\put(446,138.34){\rule{1.927pt}{0.800pt}}
\multiput(446.00,139.34)(4.000,-2.000){2}{\rule{0.964pt}{0.800pt}}
\put(454,136.34){\rule{1.686pt}{0.800pt}}
\multiput(454.00,137.34)(3.500,-2.000){2}{\rule{0.843pt}{0.800pt}}
\put(461,134.34){\rule{1.686pt}{0.800pt}}
\multiput(461.00,135.34)(3.500,-2.000){2}{\rule{0.843pt}{0.800pt}}
\put(468,131.84){\rule{1.927pt}{0.800pt}}
\multiput(468.00,133.34)(4.000,-3.000){2}{\rule{0.964pt}{0.800pt}}
\put(476,129.34){\rule{1.686pt}{0.800pt}}
\multiput(476.00,130.34)(3.500,-2.000){2}{\rule{0.843pt}{0.800pt}}
\put(483,127.84){\rule{1.927pt}{0.800pt}}
\multiput(483.00,128.34)(4.000,-1.000){2}{\rule{0.964pt}{0.800pt}}
\put(491,126.34){\rule{1.686pt}{0.800pt}}
\multiput(491.00,127.34)(3.500,-2.000){2}{\rule{0.843pt}{0.800pt}}
\put(498,124.34){\rule{1.927pt}{0.800pt}}
\multiput(498.00,125.34)(4.000,-2.000){2}{\rule{0.964pt}{0.800pt}}
\put(506,122.34){\rule{1.686pt}{0.800pt}}
\multiput(506.00,123.34)(3.500,-2.000){2}{\rule{0.843pt}{0.800pt}}
\put(513,120.84){\rule{1.927pt}{0.800pt}}
\multiput(513.00,121.34)(4.000,-1.000){2}{\rule{0.964pt}{0.800pt}}
\put(521,119.34){\rule{1.686pt}{0.800pt}}
\multiput(521.00,120.34)(3.500,-2.000){2}{\rule{0.843pt}{0.800pt}}
\put(528,117.84){\rule{1.927pt}{0.800pt}}
\multiput(528.00,118.34)(4.000,-1.000){2}{\rule{0.964pt}{0.800pt}}
\put(536,116.34){\rule{1.686pt}{0.800pt}}
\multiput(536.00,117.34)(3.500,-2.000){2}{\rule{0.843pt}{0.800pt}}
\put(543,114.84){\rule{1.927pt}{0.800pt}}
\multiput(543.00,115.34)(4.000,-1.000){2}{\rule{0.964pt}{0.800pt}}
\put(551,113.84){\rule{1.686pt}{0.800pt}}
\multiput(551.00,114.34)(3.500,-1.000){2}{\rule{0.843pt}{0.800pt}}
\put(558,112.34){\rule{1.686pt}{0.800pt}}
\multiput(558.00,113.34)(3.500,-2.000){2}{\rule{0.843pt}{0.800pt}}
\put(565,110.84){\rule{1.927pt}{0.800pt}}
\multiput(565.00,111.34)(4.000,-1.000){2}{\rule{0.964pt}{0.800pt}}
\put(573,109.84){\rule{1.686pt}{0.800pt}}
\multiput(573.00,110.34)(3.500,-1.000){2}{\rule{0.843pt}{0.800pt}}
\put(580,108.84){\rule{1.927pt}{0.800pt}}
\multiput(580.00,109.34)(4.000,-1.000){2}{\rule{0.964pt}{0.800pt}}
\put(588,107.84){\rule{1.686pt}{0.800pt}}
\multiput(588.00,108.34)(3.500,-1.000){2}{\rule{0.843pt}{0.800pt}}
\put(595,106.84){\rule{1.927pt}{0.800pt}}
\multiput(595.00,107.34)(4.000,-1.000){2}{\rule{0.964pt}{0.800pt}}
\put(603,105.84){\rule{1.686pt}{0.800pt}}
\multiput(603.00,106.34)(3.500,-1.000){2}{\rule{0.843pt}{0.800pt}}
\put(610,104.84){\rule{1.927pt}{0.800pt}}
\multiput(610.00,105.34)(4.000,-1.000){2}{\rule{0.964pt}{0.800pt}}
\put(618,103.84){\rule{1.686pt}{0.800pt}}
\multiput(618.00,104.34)(3.500,-1.000){2}{\rule{0.843pt}{0.800pt}}
\put(625,102.84){\rule{1.927pt}{0.800pt}}
\multiput(625.00,103.34)(4.000,-1.000){2}{\rule{0.964pt}{0.800pt}}
\put(633,101.84){\rule{1.686pt}{0.800pt}}
\multiput(633.00,102.34)(3.500,-1.000){2}{\rule{0.843pt}{0.800pt}}
\put(640,100.84){\rule{1.927pt}{0.800pt}}
\multiput(640.00,101.34)(4.000,-1.000){2}{\rule{0.964pt}{0.800pt}}
\put(648,99.84){\rule{1.686pt}{0.800pt}}
\multiput(648.00,100.34)(3.500,-1.000){2}{\rule{0.843pt}{0.800pt}}
\put(655,98.84){\rule{1.927pt}{0.800pt}}
\multiput(655.00,99.34)(4.000,-1.000){2}{\rule{0.964pt}{0.800pt}}
\put(670,97.84){\rule{1.686pt}{0.800pt}}
\multiput(670.00,98.34)(3.500,-1.000){2}{\rule{0.843pt}{0.800pt}}
\put(677,96.84){\rule{1.927pt}{0.800pt}}
\multiput(677.00,97.34)(4.000,-1.000){2}{\rule{0.964pt}{0.800pt}}
\put(685,95.84){\rule{1.686pt}{0.800pt}}
\multiput(685.00,96.34)(3.500,-1.000){2}{\rule{0.843pt}{0.800pt}}
\put(663.0,100.0){\rule[-0.400pt]{1.686pt}{0.800pt}}
\put(700,94.84){\rule{1.686pt}{0.800pt}}
\multiput(700.00,95.34)(3.500,-1.000){2}{\rule{0.843pt}{0.800pt}}
\put(692.0,97.0){\rule[-0.400pt]{1.927pt}{0.800pt}}
\put(715,93.84){\rule{1.686pt}{0.800pt}}
\multiput(715.00,94.34)(3.500,-1.000){2}{\rule{0.843pt}{0.800pt}}
\put(707.0,96.0){\rule[-0.400pt]{1.927pt}{0.800pt}}
\put(730,92.84){\rule{1.686pt}{0.800pt}}
\multiput(730.00,93.34)(3.500,-1.000){2}{\rule{0.843pt}{0.800pt}}
\put(722.0,95.0){\rule[-0.400pt]{1.927pt}{0.800pt}}
\put(745,91.84){\rule{1.686pt}{0.800pt}}
\multiput(745.00,92.34)(3.500,-1.000){2}{\rule{0.843pt}{0.800pt}}
\put(737.0,94.0){\rule[-0.400pt]{1.927pt}{0.800pt}}
\put(760,90.84){\rule{1.686pt}{0.800pt}}
\multiput(760.00,91.34)(3.500,-1.000){2}{\rule{0.843pt}{0.800pt}}
\put(752.0,93.0){\rule[-0.400pt]{1.927pt}{0.800pt}}
\put(782,89.84){\rule{1.686pt}{0.800pt}}
\multiput(782.00,90.34)(3.500,-1.000){2}{\rule{0.843pt}{0.800pt}}
\put(767.0,92.0){\rule[-0.400pt]{3.613pt}{0.800pt}}
\put(804,88.84){\rule{1.927pt}{0.800pt}}
\multiput(804.00,89.34)(4.000,-1.000){2}{\rule{0.964pt}{0.800pt}}
\put(789.0,91.0){\rule[-0.400pt]{3.613pt}{0.800pt}}
\put(827,87.84){\rule{1.686pt}{0.800pt}}
\multiput(827.00,88.34)(3.500,-1.000){2}{\rule{0.843pt}{0.800pt}}
\put(812.0,90.0){\rule[-0.400pt]{3.613pt}{0.800pt}}
\put(842,86.84){\rule{1.686pt}{0.800pt}}
\multiput(842.00,87.34)(3.500,-1.000){2}{\rule{0.843pt}{0.800pt}}
\put(834.0,89.0){\rule[-0.400pt]{1.927pt}{0.800pt}}
\put(857,85.84){\rule{1.686pt}{0.800pt}}
\multiput(857.00,86.34)(3.500,-1.000){2}{\rule{0.843pt}{0.800pt}}
\put(849.0,88.0){\rule[-0.400pt]{1.927pt}{0.800pt}}
\put(872,84.84){\rule{1.686pt}{0.800pt}}
\multiput(872.00,85.34)(3.500,-1.000){2}{\rule{0.843pt}{0.800pt}}
\put(864.0,87.0){\rule[-0.400pt]{1.927pt}{0.800pt}}
\end{picture}
{\small FIG.\ 2: Comparison of $D^*$ for hard spheres
from the numerical data of AGW and 
EW with those from (2) (Dz.), in the range of
$S_{ex}$ as considered by Dzugutov. Here the real $S_{ex}$ values
are used, instead of the approximation of Dzugutov.}
\end{figure}

\begin{figure}
\setlength{\unitlength}{0.240900pt}
\ifx\plotpoint\undefined\newsavebox{\plotpoint}\fi
\sbox{\plotpoint}{\rule[-0.200pt]{0.400pt}{0.400pt}}%
\begin{picture}(900,594)(0,0)
\font\gnuplot=cmr10 at 10pt
\gnuplot
\sbox{\plotpoint}{\rule[-0.200pt]{0.400pt}{0.400pt}}%
\put(120.0,82.0){\rule[-0.200pt]{4.818pt}{0.400pt}}
\put(100,82){\makebox(0,0)[r]{0.0}}
\put(859.0,82.0){\rule[-0.200pt]{4.818pt}{0.400pt}}
\put(120.0,271.0){\rule[-0.200pt]{4.818pt}{0.400pt}}
\put(100,271){\makebox(0,0)[r]{0.4}}
\put(859.0,271.0){\rule[-0.200pt]{4.818pt}{0.400pt}}
\put(120.0,460.0){\rule[-0.200pt]{4.818pt}{0.400pt}}
\put(100,460){\makebox(0,0)[r]{0.8}}
\put(859.0,460.0){\rule[-0.200pt]{4.818pt}{0.400pt}}
\put(120.0,82.0){\rule[-0.200pt]{0.400pt}{4.818pt}}
\put(120,41){\makebox(0,0){0}}
\put(120.0,534.0){\rule[-0.200pt]{0.400pt}{4.818pt}}
\put(396.0,82.0){\rule[-0.200pt]{0.400pt}{4.818pt}}
\put(396,41){\makebox(0,0){0.4}}
\put(396.0,534.0){\rule[-0.200pt]{0.400pt}{4.818pt}}
\put(672.0,82.0){\rule[-0.200pt]{0.400pt}{4.818pt}}
\put(672,41){\makebox(0,0){0.8}}
\put(672.0,534.0){\rule[-0.200pt]{0.400pt}{4.818pt}}
\put(120.0,82.0){\rule[-0.200pt]{182.843pt}{0.400pt}}
\put(879.0,82.0){\rule[-0.200pt]{0.400pt}{113.705pt}}
\put(120.0,554.0){\rule[-0.200pt]{182.843pt}{0.400pt}}
\put(798,508){\makebox(0,0){$D/D_B$}}
\put(539,21){\makebox(0,0){$n \sigma^3$}}
\put(182,389){\makebox(0,0)[l]{AGW,EW}}
\put(417,200){\makebox(0,0)[l]{Dz.}}
\put(120.0,82.0){\rule[-0.200pt]{0.400pt}{113.705pt}}
\put(130,545){\raisebox{-.8pt}{\makebox(0,0){$\Diamond$}}}
\put(159,526){\raisebox{-.8pt}{\makebox(0,0){$\Diamond$}}}
\put(169,520){\raisebox{-.8pt}{\makebox(0,0){$\Diamond$}}}
\put(174,516){\raisebox{-.8pt}{\makebox(0,0){$\Diamond$}}}
\put(218,493){\raisebox{-.8pt}{\makebox(0,0){$\Diamond$}}}
\put(315,441){\raisebox{-.8pt}{\makebox(0,0){$\Diamond$}}}
\put(364,432){\raisebox{-.8pt}{\makebox(0,0){$\Diamond$}}}
\put(446,385){\raisebox{-.8pt}{\makebox(0,0){$\Diamond$}}}
\put(608,263){\raisebox{-.8pt}{\makebox(0,0){$\Diamond$}}}
\put(662,219){\raisebox{-.8pt}{\makebox(0,0){$\Diamond$}}}
\put(695,192){\raisebox{-.8pt}{\makebox(0,0){$\Diamond$}}}
\put(730,162){\raisebox{-.8pt}{\makebox(0,0){$\Diamond$}}}
\put(771,129){\raisebox{-.8pt}{\makebox(0,0){$\Diamond$}}}
\sbox{\plotpoint}{\rule[-0.500pt]{1.000pt}{1.000pt}}%
\put(120,82){\usebox{\plotpoint}}
\put(120.00,82.00){\usebox{\plotpoint}}
\put(140.71,82.71){\usebox{\plotpoint}}
\put(161.35,84.42){\usebox{\plotpoint}}
\put(181.72,88.09){\usebox{\plotpoint}}
\put(202.15,91.47){\usebox{\plotpoint}}
\put(222.25,96.64){\usebox{\plotpoint}}
\put(242.35,101.84){\usebox{\plotpoint}}
\put(262.13,108.03){\usebox{\plotpoint}}
\put(281.85,114.32){\usebox{\plotpoint}}
\put(301.34,121.33){\usebox{\plotpoint}}
\put(320.79,128.45){\usebox{\plotpoint}}
\put(340.34,135.29){\usebox{\plotpoint}}
\put(360.08,141.59){\usebox{\plotpoint}}
\put(379.88,147.72){\usebox{\plotpoint}}
\put(399.63,153.91){\usebox{\plotpoint}}
\put(419.38,160.09){\usebox{\plotpoint}}
\put(439.58,164.70){\usebox{\plotpoint}}
\put(459.91,168.73){\usebox{\plotpoint}}
\put(480.37,172.05){\usebox{\plotpoint}}
\put(501.00,174.00){\usebox{\plotpoint}}
\put(521.70,175.00){\usebox{\plotpoint}}
\put(542.45,174.94){\usebox{\plotpoint}}
\put(563.09,173.24){\usebox{\plotpoint}}
\put(583.67,170.54){\usebox{\plotpoint}}
\put(604.01,166.75){\usebox{\plotpoint}}
\put(624.09,161.48){\usebox{\plotpoint}}
\put(644.16,156.21){\usebox{\plotpoint}}
\put(663.95,150.02){\usebox{\plotpoint}}
\put(683.59,143.46){\usebox{\plotpoint}}
\put(703.24,136.90){\usebox{\plotpoint}}
\put(722.54,129.30){\usebox{\plotpoint}}
\put(742.13,122.58){\usebox{\plotpoint}}
\put(761.63,115.59){\usebox{\plotpoint}}
\put(781.41,109.40){\usebox{\plotpoint}}
\put(801.20,103.23){\usebox{\plotpoint}}
\put(821.30,98.06){\usebox{\plotpoint}}
\put(841.59,93.92){\usebox{\plotpoint}}
\put(861.99,90.25){\usebox{\plotpoint}}
\put(879,88){\usebox{\plotpoint}}
\sbox{\plotpoint}{\rule[-0.200pt]{0.400pt}{0.400pt}}%
\put(396,82){\usebox{\plotpoint}}
\put(396.0,82.0){\rule[-0.200pt]{0.400pt}{113.705pt}}
\put(706,82){\usebox{\plotpoint}}
\put(706.0,82.0){\rule[-0.200pt]{0.400pt}{113.705pt}}
\sbox{\plotpoint}{\rule[-0.400pt]{0.800pt}{0.800pt}}%
\put(396,153){\usebox{\plotpoint}}
\multiput(396.00,154.40)(1.943,0.520){9}{\rule{3.000pt}{0.125pt}}
\multiput(396.00,151.34)(21.773,8.000){2}{\rule{1.500pt}{0.800pt}}
\multiput(424.00,162.39)(2.806,0.536){5}{\rule{3.800pt}{0.129pt}}
\multiput(424.00,159.34)(19.113,6.000){2}{\rule{1.900pt}{0.800pt}}
\multiput(451.00,168.38)(4.286,0.560){3}{\rule{4.680pt}{0.135pt}}
\multiput(451.00,165.34)(18.286,5.000){2}{\rule{2.340pt}{0.800pt}}
\put(479,171.34){\rule{6.504pt}{0.800pt}}
\multiput(479.00,170.34)(13.500,2.000){2}{\rule{3.252pt}{0.800pt}}
\put(506,172.84){\rule{6.745pt}{0.800pt}}
\multiput(506.00,172.34)(14.000,1.000){2}{\rule{3.373pt}{0.800pt}}
\put(534,172.34){\rule{6.745pt}{0.800pt}}
\multiput(534.00,173.34)(14.000,-2.000){2}{\rule{3.373pt}{0.800pt}}
\put(562,169.34){\rule{5.600pt}{0.800pt}}
\multiput(562.00,171.34)(15.377,-4.000){2}{\rule{2.800pt}{0.800pt}}
\multiput(589.00,167.06)(4.286,-0.560){3}{\rule{4.680pt}{0.135pt}}
\multiput(589.00,167.34)(18.286,-5.000){2}{\rule{2.340pt}{0.800pt}}
\multiput(617.00,162.08)(1.870,-0.520){9}{\rule{2.900pt}{0.125pt}}
\multiput(617.00,162.34)(20.981,-8.000){2}{\rule{1.450pt}{0.800pt}}
\multiput(644.00,154.08)(1.684,-0.516){11}{\rule{2.689pt}{0.124pt}}
\multiput(644.00,154.34)(22.419,-9.000){2}{\rule{1.344pt}{0.800pt}}
\multiput(672.00,145.08)(1.684,-0.516){11}{\rule{2.689pt}{0.124pt}}
\multiput(672.00,145.34)(22.419,-9.000){2}{\rule{1.344pt}{0.800pt}}
\end{picture}
{\small FIG.\ 3: Comparison between the numerical data of AGW and EW for
$D/D_B$ for hard spheres and those obtained from (2).
The the vertical lines indicate the range considered by Dzugutov.}
\end{figure}

\noindent
E.G.D.\ Cohen, Rockfeller University, New York, NY 10021 - U.S.A.

\noindent
L.\ Rondoni, 
Politecnico di Torino, C. D. Abruzzi 24, I-10129 Torino - Italy

\vskip 10pt
\noindent
PACS numbers: 05.70Ce, 05.45.+b, 61.20.Ja
\vskip 10pt
\noindent
[1] M.\ Dzugutov, et al.\, Phys. Rev. Lett.\underline{81},
1762 (1998)

\noindent
[2] Ch.\ Dellago and H.A.\ Posch, Physica A, {\bf 240}, 68 (1997)

\noindent
[3] M.\ Dzugutov, Nature, {\bf 381}, 137 (1996)

\noindent
[4] B.J.\ Alder, et al.\, J.\ Chem.\ Phys.\,
{\bf 53}, 3813 (1970)

\noindent
[5] J.J.\ Erpenbeck et al.\, Phys.\ Rev.\ A {\bf 43}, 4254 (1991)

\end{multicols}

\end{document}